\documentclass[12pt]{article}

\def\rdots{\mathinner{\mkern1mu\raise1pt\vbox{\kern1pt\hbox{.}}\mkern2mu
   \raise4pt\hbox{.}\mkern2mu\raise7pt\hbox{.}\mkern1mu}}
\newcommand{\be}{\begin{equation}}
\newcommand{\ee}{\end{equation}}
\newcommand{\Z}{{\rm Z\kern-.35em Z}}
\newcommand{\bP}{{\rm I\kern-.15em P}}
\newcommand{\Q}{\kern.3em\rule{.07em}{.65em}\kern-.3em{\rm Q}}
\newcommand{\R}{{\rm I\kern-.15em R}}
\newcommand{\h}{{\rm I\kern-.15em H}}
\newcommand{\C}{\kern.3em\rule{.07em}{.65em}\kern-.3em{\rm C}}
\newcommand{\T}{{\rm T\kern-.35em T}}
\newcommand{\D}{{\kern-.5em /}}

\begin{document}

\openup 1.5\jot

\centerline{$e$ to the $A$, in a New Way}

\vspace{1in}
\centerline{Paul Federbush}
\centerline{Department of Mathematics}
\centerline{University of Michigan}
\centerline{Ann Arbor, MI 48109-1109}
\centerline{(pfed@math.lsa.umich.edu)}

\vspace{1in}

\centerline{Abstract}

Apparently new expressions are given for the exponential of a hermitian matrix, $A$, in the $2 \times 2, 3 \times 3$, and $4 \times 4$ cases.  Replacing $A$ by $iA$ these are explicit formulas for the Fourier transform of $e^{iA}$.

\vspace{.50in}

The support of the Fourier transform of $e^{iA}$, $A$ hermitian, has been long known [1], [2], [3] and [4].  We exhibit explicit expressions for traceless hermitian matrices of sizes $2 \times 2, 3 \times 3$, and $4 \times 4$.  In $d$-dimensions, where $A
$ acts on $\C^d$, we employ matrices, $W$, hermitian projections of rank one.  Thus $W$ may be written as $W_{ij} = v_i \; \bar{v}_j$, where $v_i$ is a unit vector in $\C^d$, and Tr$(AW) = < v, Av >$.  $\int d \Omega$ denotes a normalized integral over al
l such $W$ defined by integrating the associated $v_i$ over a unit sphere in $\C^d$ with the normalized unitary- invariant measure.  The expressions we give are not unique, among similar forms.  It is clear one can derive such expressions in any number of
 dimensions.  We leave to others the task of creating an appropriate theory.

$2-d$ case
\be
e^A = \int d\Omega e^{{\rm Tr}(AW)} \left[ I + 4( W - \frac 1 2 I) + \ {\rm Tr}(AW)I + 2 \ {\rm Tr}(AW) ( W - \frac 1 2 I) \right]
\ee

$3-d$ case
\be
e^A = \int d\Omega e^{{\rm Tr}(AW)} \Bigg[ I + 9( W - \frac 1 3 I) - \ {\rm Tr}(AW)I + 9 \ {\rm Tr}(AW) W
\ee
\[
- \frac 1 2 \; A + \left[ \frac 3 2 ( {\rm Tr}(AW))^2 - \frac 1 4 \; \ {\rm Tr}(A^2) \right] W \Bigg].
\]

$4-d$ case
\be
e^A = \int d\Omega e^{{\rm Tr}(AW)} \Bigg[ I + \frac {52}{3}( W - \frac 1 4 I)  - \frac{8}{3}   \ {\rm Tr}(AW)I + \frac{68}{3}  \ {\rm Tr}(AW) W-A
\ee
\[
+ \frac 1 6 \left[ - 4\ {\rm Tr}(A^2) + 46 ( {\rm Tr}(AW))^2 + 2A\ {\rm Tr}(AW)\right] W
\]
\[
- \frac 1 6 \left[ - \frac 1 2 \ {\rm Tr}(A^2)I + 3( {\rm Tr}(AW))^2 I + A^2 + 2A \ {\rm Tr}(AW) \right]
\]
\[
+ \frac 2 3 ( {\rm Tr}(AW))^3 W - \frac {1}{18} \ {\rm Tr}(A^3)W - \frac 1 6 \ {\rm Tr}(A^2) \ {\rm Tr}(AW)W \Bigg].
\]

These expressions were derived using such trickery and chicanery as will likewise not be useful to one proving a general theory:  extracting leading and subleading asymptotic behaviors as one eigenvalue of $A$ approaches $\infty$ (see Appendix B), finding
 the simple rational numbers involved in the ``angular integrals" of $W$ by numerical integration (see Appendix A).  The expressions were likewise checked by numerical integration.  The $2-d$ expression of eq. (1) is intimately related to the expression d
erived in [4] for this dimension.

\vfill\eject

\centerline{\underline{Appendix A}}

In this Appendix we collect the most useful of the ``angular integrals" computed.  All of these results were derived numerically, but analytic derivation should be straightforward, if tedious.  We denote

\be	Av(f) \equiv \int d \Omega f		\ee
for any function of $W, f$.

2-d
\be	Av(Tr(AW)W) = \frac 1 6 \ A		\ee

3-d

\be 	Av(Tr(AW)W) = \frac 1 {12} \ A		\ee
\be 	Av \Big( (Tr(AW))^2 W \Big) = \frac 1 {30} \ A^2 + \frac 1 {60} Tr (A^2)	\ee
\be 	Av \Big( (Tr(AW))^3 W \Big) = \frac 1 {30} \ A^3 	\ee

4-d

\be 	Av(Tr(AW)W) = \frac 1 {20} \ A		\ee
\be 	Av \Big( (Tr(AW))^2 W \Big) = \frac 1 {120} \ Tr(A^2) + \frac 1 {60}  A^2	 \ee
\be 	Av \Big( (Tr(AW))^3 W \Big) = \frac 1 {420} \ Tr(A^3) + \frac 1 {280} \ Tr(A^2)A 
+ \frac 1 {140} \ A^3.		\ee

\vfill\eject
\centerline{\underline{Appendix B}}

In this Appendix we sketch the computations we used to derive the formula, eq. (2), in 3 dimensions.  The derivation of the 4 dimensional case, eq. (3), is similar but more complex.  We let $A$ have the form
\be
\left( \begin{array}{ccc}
1 & 0 & 0 \\
0 & \lambda_1 & 0 \\
0 & 0 & \lambda_2
\end{array}   \right)
\ee
with
\be     1+ \lambda_1 + \lambda_2 = 0	\ee
and
\be	1 > \lambda_1, \ \ \ 1 > \lambda_2 \ .	\ee
We also set
\be
W_0 = \left( \begin{array}{ccc}
1 & 0 & 0 \\
0 & 0& 0 \\
0 & 0 & 0
\end{array}   \right) .
\ee
We then find (ala steepest descent, or Laplace's Method) that
\be	I(s) \equiv \int d\Omega e^{s\; Tr(AW)} \; P(W)	\ee
as $s$ goes to $+\infty$, has leading asymptotic behavior coming entirely from an arbitrarily small neighborhood of $W_0$ and
\be	I(s) \sim \frac 2 {s^2(1-\lambda_1)(1-\lambda_2)} \ e^s P(W_0).    \ee
A little calculation shows
\be     (1-\lambda_1)(1-\lambda_2) = 3 \left( Tr(AW_0) \right)^2 - \frac 1 2 \; Tr(A^2). \ee
Equations (16)-(18) explain the choice of terms in (2) quadratic in $A$.  We found terms linear in $A$ by writing the most general invariant linear expression and checking the supposed identity eq. (2) order by order in $A$, using the formulae in Appendix
 A.  (If we allowed the invariant $Tr(A^2W)$ to appear in the quadratic terms, say by adding the expression $Tr(A^2 W) - (Tr(AW))^2$ to the quadratic terms, and appropriately modifying the linear terms,  alternate expressions for the identity eq. (2) were
 obtained.)

\vfill\eject

\centerline{REFERENCES}
\begin{description}
\item[[1]]  E. Nelson, {\it Operants: A functional calculus for non-commuting operators}, Functional Analysis and Related Fields, Proceedings of a conference in honor of Professor Marshal Stone (Univ. of Chicago, May 1968) (F.E. Browder, ed.), Springer-Ve
rlag, Berlin, Heidelberg, and New York, 1970, pp. 172-187.  MR 54:978.
\item[[2]]  B. Jefferies, ``The Weyl Calculus for Hermitian Matrices", {\it Proc. A.M.S.} {\bf 124} (96) p. 121-128.
\item[[3]] M.E. Taylor ``Functions of Several Self-Adjoint Operators", {\it Proc. A.M.S.} {\bf 19} (1968), 91-98.  MR {\bf 36}:3149.
\item[[4]] R.F.V. Anderson, ``The Weyl Functional Calculus", {\it J. Func. Anal.} {\bf 4} (1969) 240-267.  MR {\bf 58}:30405.

\end{description}

\end{document}